\documentclass{article}
\usepackage{ijcai24}

\usepackage{times}
\usepackage{soul}
\usepackage{url}
\usepackage[hidelinks]{hyperref}
\usepackage[utf8]{inputenc}
\usepackage[small]{caption}
\usepackage{subcaption}
\usepackage{graphicx}
\usepackage{amsmath}
\usepackage{amsthm}
\usepackage{booktabs}
\usepackage{algorithm}
\usepackage{algorithmic}
\usepackage[switch]{lineno}
\usepackage{acronym}
\usepackage[english]{babel}

\usepackage{changes}
\usepackage{paralist}
\usepackage{acronym}
\usepackage{mathrsfs}  

\usepackage{todonotes}
\usepackage{algorithm}
\usepackage{amsmath}
\usepackage{amsthm}
  \newcommand{\nat}{\mathbf{N}}

  \newcommand{\eg}{\textit{e.g.}}

\newcommand{\dist}{\mathcal{D}}

  \newcommand{\supp}{\mathsf{Supp}}

\newcommand{\reals}{\mathbf{R}}
 
\newtheorem{definition}{Definition}

\newtheorem{problem}{Problem}

\newtheorem{example}{Example}

\acrodef{mdp}[MDP]{Markov decision process} 
\acrodef{hmm}[HMM]{Hidden Markov Model}

 \DeclareMathOperator*{\optst}{\textrm{subject to}}

     \newcommand{\obs}{\mathsf{Obs}}

\usepackage{mathtools}
\usepackage{graphicx}

\usepackage{tikz}
\usetikzlibrary{arrows, arrows.meta}
\usetikzlibrary{automata,positioning}
\usetikzlibrary{shapes.geometric}

\definecolor{darkgreen}{rgb}{0,0.5,0}

\title{Information-Theoretic Opacity-Enforcement in Markov Decision Processes}

\author{
Chongyang Shi 
\and
Yuheng Bu 
\And
Jie Fu 
\affiliations
 University of Florida\\
\emails
\{c.shi, buyuheng, fujie\}@ufl.edu
}

\begin{document}

\maketitle

\begin{abstract}
The paper studies information-theoretic opacity, an information-flow privacy property, in a setting involving two agents: A planning agent who controls a stochastic system and an observer who partially observes the system states. The goal of the observer is to infer some secret, represented by a random variable, from its partial observations, while the goal of the planning agent is to make the secret maximally opaque to the observer while achieving a satisfactory total return. Modeling the stochastic system using a Markov decision process, two classes of opacity properties are considered---Last-state opacity is to ensure that the observer is uncertain if the last state is in a specific set and initial-state opacity is to ensure that the observer is unsure of the realization of the initial state. As the measure of opacity, we employ the Shannon conditional entropy capturing the information about the secret revealed by the observable. Then, we develop primal-dual policy gradient methods for opacity-enforcement planning subject to constraints on total returns. We propose novel algorithms to compute the policy gradient of entropy for each observation, leveraging message passing within the hidden Markov models. This gradient computation enables us to have stable and fast convergence. We demonstrate our solution of opacity-enforcement control through a grid world example.
\end{abstract}

\section{Introduction} 
Opacity, as a plural concept, has been proposed to generalize secrecy, anonymity, privacy, and other confidentiality properties against attacks \cite{zengQuantitativeAnalysisOpacity2021,watsonMultiLevelSecurityModel2011,guoOverviewOpacityDiscrete2020}. 
Opacity enforcement is to hide sensitive information from an external observer. 

Depending on the various measures of opacity, two classes of opacity enforcement problems have been investigated:  Qualitative opacity enforcement requires that an adversary/observer, with complete knowledge of the system yet noisy observations, can ascertain some ``secret'' information of the system with probability one. 
Quantitative opacity enforcement, on the other hand, requires that the observer's uncertainty in \emph{guessing the secret}  after receiving the observations is close to the uncertainty the observer initially had. 
Connecting with the classical work of Shannon on secrecy 
\cite{shannonCommunicationTheorySecrecy1949} and \emph{guesswork} \cite{khouzani2017leakage}, we propose to measure quantitative opacity using the conditional entropy of the random variable (e.g. the observer's estimate of a secret variable $Z$) given the observed information (e.g. the observation sequence $Y$). This is because
$\text{the leaked information} =$ 
\[
\underbrace{\text{initial uncertainty}}_{H(Z)- \text{``entropy before observation''}} - \underbrace{\text{remaining uncertainty}}_{H(Z|Y) - \text{`` entropy after observation''}}.
\]
The more opaque the system is, the larger the entropy after the observation. With this notion of quantitative opacity, we study optimal opacity-enforcement planning in a stochastic system modeled as a \ac{mdp} subject to task performance constraints. We consider the interaction between a planning agent (referred to as player 1/P1) and an observer (referred to as player 2/P2). The goal of P1 is to maximize the opacity while ensuring that the total return meets a satisfying threshold, against P2 who has partial observations about the system trajectories but complete knowledge about the system model and P1's control policy.
Two problems of state-based opacity are studied. One is the \emph{last-state} opacity, where P1 aims to obfuscate if the last state is in a set of secret states, to P2. The other one is \emph{initial-state} opacity, where P1 aims to obfuscate the \emph{exact initial state} to P2. 

\noindent \textbf{Related Work}
Opacity was originally introduced by Mazaré \shortcite{Mazar2004UsingUF}, referring to concealing secrets from an observer. This work has led to   \textit{qualitative} opacity analyses  within discrete event systems (DESs), where the system dynamics are governed by both controllable and uncontrollable events, and its information leakage is also defined by observable and non-observable events. In qualitative analysis, a trajectory is opaque if it satisfies some secret property and is indistinguishable from at least one trajectory that does not satisfy the secret. Furthermore, a system is opaque if it generates only an opaque trajectory. 
Different types of qualitative opacity have been studied, including \emph{state-based}, 
which requires the secret behavior of the system (i.e., the membership of its initial/current/past state to the set) to remain opaque   \cite{saboori2007notions,saboori2013current,HanX2023Scai,saboori_verification_2013,Saboori2009Kstep,yinInfinitestepOpacityKstep2019};  
\emph{language-based} \cite{dubreil2008opacity,lin2011opacity,shi2023synthesis}, which aims to hide a set of secret executions; and ``\emph{model-based}'' \cite{keroglouProbabilisticSystemOpacity2016}, aiming to prevent the observer from finding out the true model of the system among several candidates. In multi-agent deterministic systems, cooperative privacy-preserving planning has been investigated   \cite{Maliahmultiprivacy2017,shani2018advances} to achieve a global coordinated plan without revealing local private information. 

A \textit{quantitative} analysis of opacity is essential for measuring the degree of opacity in a system. \cite{berardProbabilisticOpacityMarkov2015} defines quantitative opacity as the probability of generating an opaque run in a stochastic system. A related notion to quantitative opacity is covertness \cite{Marzouqi2004Covert,Marzouqi2005Covert}, which can be viewed as a model-based opacity, as it makes the observer uncertain if a covert agent is present or not.
 Recent work \cite{ma2023covert} developed covert planner that leverages the coupling of stochastic dynamics and observational noises to make the covert agent's policy indistinguishable from a nominal policy.
Information-theoretic opacity has been studied in channel design problems, which can be viewed as a one-step decision-making problem\cite{khouzani2017leakage}. 
Another recent work \cite{molloy2023smoother} developed a method for obfuscating state trajectories in partially observable Markov decision processes (POMDPs) and shows that the causal conditional entropy of the state trajectory can be expressed into a cumulative sum, permitting the use of POMDP solvers. Our work is also related to observer-aware MDP \cite{pmlr-v161-miura21a} in the sense that one of the reward functions depends on the observer's belief (of the secret). However, our solution does not employ POMDP solvers.

\noindent \textbf{Our Contribution}
Our work can be seen as a generalization of minimal information leakage channel design from one-stage decision-making to sequential decision-making in Markov decision processes, concerning state-based opacity. 
We establish the last-state opacity and initial-state opacity by leveraging Shannon entropy \cite{shannonCommunicationTheorySecrecy1949} as a \emph{symmetric} measure of opacity. 
Different from Molloy and Nair \shortcite{molloy2023smoother}, we do not assume that the observer has access to control inputs. Thus, the conditional entropy of state-based opacity properties cannot be constructed as a cumulative sum of entropy terms. We propose a novel approach to compute/estimate the gradient of the conditional entropy with respect to (w.r.t.) the parameters of the control policy, leveraging the forward-backward algorithm in HMMs. Building on the gradient estimation/computation, we can then formulate the optimal opacity enforcement planning problem, subject to task constraints, into a constrained optimization and employ primal-dual gradient-based optimization to compute a (locally)-optimal policy that maximizes the opacity while satisfying a constraint on the total return. The correctness of the proposed methods is experimentally validated.




  
 
\section{Preliminary and Problem Formulation}
\label{sec: preliminary}

\subsection{Preliminaries}
\noindent \textbf{Notation} The set of real numbers is denoted by $\reals$. Random variables will be denoted by capital letters, and their realizations by lowercase letters (\eg, $X$ and $x$). The probability mass function (pmf) of a discrete random variable $X$ will be written as $p(x)$, the joint pmf of $X$ and $Y$ as $p(x, y)$, and the conditional pmf of $X$ given $Y = y$ as $p(x|y)$ or $p(x|Y = y)$. The sequence of random variables and their realizations with length $T$ are denoted as $X_{[0:T]}$ and $x_{[0:T]}$, respectively. Given a finite and discrete set $\mathcal{S}$, let $\dist(\mathcal{S})$ be the set of all probability distributions over $\mathcal{S}$. The set $\mathcal{S}^{T}$ denotes the set of sequences with length $T$ composed of elements from $\mathcal{S}$, and $\mathcal{S}^\ast$ denotes the set of all finite sequences generated from $\mathcal{S}$.


\paragraph*{The Planning Problem} The problem is modeled as an \ac{mdp} $M=\langle \mathcal{S}, \mathcal{A}, P,\mu_0, R, \gamma \rangle$ where $\mathcal{S}$ is a finite set of states, $\mathcal{A}$ is a finite set of actions, $P:\mathcal{S} \times \mathcal{A}\rightarrow \dist(\mathcal{S})$ is a probabilistic transition function and  $P(s'|s, a)$ is the probability of reaching state $s'$ given that action $a$ is taken at the state $s$, and $\mu_0$ is the initial state distribution. The planning objective is described by a reward function $R: \mathcal{S}\times \mathcal{A}\rightarrow \reals$. $\gamma \in [0,1]$ is the discount factor.


We refer to the planning agent as player 1, or P1, and the observer as player 2, or P2. 
For a Markov policy $\pi:\mathcal{S} \rightarrow \dist(\mathcal{A})$,  P1's value function $V  ^{\pi}: \mathcal{S} \rightarrow \reals$ is defined as
\[
V ^{\pi}(s) = E_{\pi}[\sum\limits_{k = 0}^{\infty}\gamma^{k}R(S_k, \pi(S_k))|S_0 =s],
\] where $E_{\pi}$ is the expectation w.r.t. the probability distribution induced by the policy $\pi$ from $M$. We denote the Markov chain induced by the policy $\pi$ as $M_\pi$. And $S_k$ is the $k$-th state in the Markov chain $M_\pi$. 
P2's observation function is a common knowledge, defined as follows:

\begin{definition}[Observation function of P2]
Let $\mathcal{O}$ be a finite set of observations. The state-observation function of P2 is $\obs: \mathcal{S}  \rightarrow \dist(\mathcal{O})$ that maps a state $s$ to a distribution $\obs(s)$ in the observation space. The action is non-observable.
\end{definition}
Note that the assumption on non-observable actions can be easily relaxed, by augmenting the state space with $\mathcal{S}\cup (\mathcal{S}\times \mathcal{A})$ and defining the state-observation function over the augmented state space. This assumption is made for clarity.

For an MDP $M$ and P2's observation function $\obs$, a policy $\pi$ induces a discrete stochastic process $\{S_t, O_t, t\in \nat\}$ where each $S_t \in \mathcal{S}$ and $O_t \in \mathcal{O}$. For a Markov policy, 
\[
P(S_{t+1}=s' | S_t = s) = \sum_{a\in \mathcal{A}} P(s'|s,a)\pi(a|s),
\]
and $P(O_t = o | S_t=s) =  \obs(o| s ). $

Next, we will introduce some basic definitions in information theory that are used to quantify the opacity of a policy. 

The \emph{entropy of a random variable} $X$ with a countable support $\cal X$ and a probability mass function $p$ is 
\[
H(X) = -\sum_{x\in \mathcal{X}}p(x)\log p(x).
\]
The joint entropy of two random variables $X_1,X_2$ with the same support $\mathcal{X}$ is 
\[
H(X_1,X_2) =  -\sum_{x_1\in \mathcal{X}}\sum_{x_2\in \mathcal{X}} p(x_1,x_2)\log p(x_1,x_2).
\]

The conditional entropy of $X_2$ given $X_1$ is 
\[
H(X_2|X_1) =   -\sum_{x_1\in \mathcal{X}}\sum_{x_2\in \mathcal{X}} p(x_1,x_2) \log p(x_2|x_1).
\]
The conditional entropy measures the uncertainty about $X_2$ given knowledge of $X_1$.  It is defined as 
$$
H(X_2|X_1) = H(X_1,X_2) - H(X_1).
$$
A higher conditional entropy makes it more challenging to infer $X_2$ from $X_1$.


\subsection{Problem Statement}

\begin{definition} 
Given the \ac{mdp} $M$ and a policy $\pi$ induce a discrete stochastic process $M_\pi = \{S_t, O_t, t\in \nat\}$ with $T>0$ a finite number. Let $E$ be a set of \emph{secret} states. For any $0\le t\le T$, let the random variable $Z_t$ be defined by 
\[ Z_t = \mathbf{1}_E (S_t).\] 
That is, $Z_t$ is the random variable representing if the $t$-th state is in the set of secret states. The random variables $Z_t$ are binary $\mathcal{Z} = \{0,1\}$, for each $0 \le t \le T$.
\end{definition}

The conditional entropy of  $Z_t$ given an observation $Y= O_{[0:T]}$ is defined as  
\begin{multline}
H(Z_t|Y; {M_\pi}) =
-\sum_{z \in \mathcal{Z}}\sum_{y \in \mathcal{O}^T} P^{M_\pi}(Z_t= z, Y= y) \\ \cdot \log_2 P^{M_\pi}(z|Y=y).
\end{multline}
where $P^{M_\pi}(Z_t=z,y)$ is the joint probability of $Z_t=z$ (for $z=0,1$) and observation $y$ given the stochastic process $M_\pi$ and  $P^{M_\pi}(z|Y=y)$ is the conditional probability of $z$ given the observation $y$ and $\mathcal{O}^T$ is the sample space for the observation sequence of length $T$. This conditional entropy \cite{shannonCommunicationTheorySecrecy1949} can be interpreted as the fewest number of subset-membership queries an adversary must make before discovering the secret.

Based on the notion of information leakage  \cite{yasuokaQuantitativeInformationFlow2010},
the quantitative analysis of confidential information of a dynamical system is defined as the difference between an attacker's \emph{capability in guessing the secret} before and after available observations about the system. Thus, the maximal opacity-enforcement planning with task constraints can be formulated as the following problem:
\begin{problem}[Maximal last-state opacity]
    Given the \ac{mdp} $M$, a set $E$ of secret states, a finite horizon $T$,  compute a policy that maximizes the conditional entropy $H(Z_T|Y; M_\pi)$ between the random variable $Z_T$ and the observation sequence $Y = O_{[0:T]}$ while ensuring the total discounted reward exceeds a given threshold $\delta$.
\end{problem}
In other words, the objective is to ensure the adversary, with the knowledge of the agent's policy and the observations, is maximally uncertain regarding whether the last state of a finite path is in the secret set. This symmetric notion of opacity implies that opacity will be minimal when the observer is consistently confident that the agent either visited or avoided the secret states given an observation. This stands in contrast to asymmetric opacity \cite{berardQuantifyingOpacity2015}, which only assesses the uncertainty regarding the visits of secrets. Meanwhile, the agent is required to get an adequate total reward (no less than $\delta$)  to achieve a satisfactory task performance.

We also consider the problem of hiding the initial state. 
We assume that the observer has the complete knowledge of MDP, and thereby knows the initial state distribution $\mu_0$. Then, the nature selects one of the initial states probabilistically according to $\mu_0$, says $s_0$. The agent aims to hide the actual realization of the initial state $s_0$ from the observer.
\begin{problem}[Maximal initial-state opacity]
\label{problem:initial-state}
    Given the \ac{mdp} $M$, a finite horizon $T$, and the initial state distribution $\mu_0$.
    Compute a policy that maximizes the conditional entropy $H(S_0  |Y; {M_\pi})$ between the initial state $S_0$ and the observation sequence $Y = O_{[0:T]}$ while ensuring the total discounted reward exceeds a given threshold $\delta$.
\end{problem}


\section{Synthesizing Maximally Opacity-Enforcement Controllers For Last-state Opacity}

\subsection{Primal-Dual Policy Gradient for Constrained Minimal Information Leakage}

For clarity in notation, we consider a finite state set $\mathcal{S}=\{1,\ldots, N\}$. 
We introduce a set of parameterized Markov policies $\{\pi_\theta\mid \theta\in \Theta\}$ where $\Theta$ is a finite-dimensional parameter space. 
For any Markov policy $\pi_\theta$ parameterized by  $\theta$, the Markov chain induced by $\pi_\theta $ from $M$ is denoted $M_\theta \colon \{S_t, A_t, t\ge 0\}$ where $S_t$ is the random variable for the $t$-th state and $A_t$ is the random variable for the $t$-th action. The transition kernel of the Markov chain is $P_\theta$ such that $P_\theta(i,j)  =  \sum_{a\in \mathcal{A}} P(j|i,a)\pi_\theta(a|i)$ is the probability of moving from state $i$ to $j$ in one step. 

The maximal last-state opacity-enforcement planning problem can be formulated as a constrained optimization problem as follows:
\begin{equation}
\begin{aligned}
\label{eq:opt_problem_1}
& \max_\theta && H (Z_T|Y;\theta) \\
&\optst: &&  V(\mu_0,\theta) \geq \delta,
\end{aligned}
\end{equation}
where $\delta$ is a lower bound on the value function. The value $V(\mu_0,\theta)$ is obtained by evaluating the policy $\pi_\theta$ given the initial state distribution $\mu_0$, i.e., $V(\mu_0, \theta) \coloneqq V^{\pi_\theta}(\mu_0)$, and $H(Z_T|Y;\theta)  \coloneqq H(Z_T|Y;  M_\theta)$.


For this constrained optimization problem with an inequality constraint,   we formulate the problem \eqref{eq:opt_problem_1} into the following max-min problem for the associated Lagrangian $L (\theta, \lambda)$:
\begin{equation}
\label{eq:opt_problem_2}
\max_\theta \min_{\lambda \geq 0} L(\theta, \lambda) = H(Z_T|Y;\theta) + \lambda (V(\mu_0, \theta) - \delta),
\end{equation}
where $\lambda$ is the multiplier. 

In each iteration $k$, the primal-dual gradient descent-ascent algorithm is given as
\begin{equation}
\begin{aligned}
\label{eq:primal_dual_algorithm}
& \theta_{k + 1} = \theta_k + \eta \nabla_\theta L(\theta, \lambda), \\
& \lambda_{k + 1} = \lambda_k - \kappa (V(\mu_0, \theta) - \delta),
\end{aligned}
\end{equation}
where $\eta > 0, \kappa > 0$ are step sizes. And the gradient of Lagrangian function w.r.t. $\theta$ can be computed as
\begin{equation}
\nabla_\theta L(\theta, \lambda) = \nabla_\theta H(Z_T|Y;\theta) + \lambda \nabla_\theta V(\mu_0, \theta).
\end{equation}
The term $\nabla_\theta V(\mu_0, \theta)$ can be easily computed by the standard policy gradient algorithm \cite{Sutton1999policy}. 

Following the primal-dual gradient descent-ascent method, we need to compute the gradient of the conditional entropy w.r.t. the policy parameter $\theta$. This is nontrivial because this conditional entropy is non-cumulative. Next, we show how to employ the forward algorithm in \ac{hmm} to compute such a gradient $\nabla_\theta H(Z_T|Y;\theta)$.

\subsection{Computing the Gradient of Conditional Entropy}
\label{subsec:forward_gradient}

From the observer's perspective, the stochastic process under policy $\pi_\theta$ is a \ac{hmm} $HM_\theta = (\mathcal{S}, \mathcal{O}, P_\theta, B)$, where $\mathcal{S}$ is the state space, $\mathcal{O}$ is the observation state space, $P_\theta$  is the transition kernel and $B  = \{ b_{i}, i \in \mathcal{S}\}$ is the emission probability distributions where $b_{i}(o)= \obs(o|i)$.  It is noted that the emission distributions do not depend on the policy.

The conditional entropy of  $Z_T$ given an observation sequence $Y$ can be written as
\begin{equation}
\label{eq:HMM_entropy}
H(Z_T|Y;\theta) = - \sum_{y \in \mathcal{O}^T} \sum_{z_T \in \{0,1\}} P_\theta(z_T, y) \log_2 P_\theta(z_T | y), 
\end{equation}
where the probability measure $P_\theta \coloneqq P^{\mathcal{M}_{\pi_\theta}}$. The discrete conditional entropy of a binary random variable has a property that $0 \leq H(Z_T|Y;\theta) \leq 1$. We take the gradient of $H(Z_T|Y;\theta)$ w.r.t. the policy parameter $\theta$. By using a trick that $\nabla_\theta P(y) = P_\theta(y) \nabla_\theta \log P_\theta(y)$ and the property of conditional probability, we have
\begin{equation}
\begin{aligned}
\label{eq:HMM_gradient_entropy}
 &\nabla_\theta H(Z_T|Y;\theta) \\
= & - \sum_{y \in \mathcal{O}^T} \sum_{z_T \in \{0,1\}} \Big[\nabla_\theta P_\theta(z_T, y) \log_2 P_\theta(z_T | y) \\
&+  P_\theta(z_T, y) \nabla_\theta  \log_2 P_\theta(z_T | y)\Big] \\
= & - \sum_{y \in \mathcal{O}^T} \sum_{z_T \in \{0,1\}} \Big[P_\theta(y) \nabla_\theta P_\theta(z_T| y) \log_2 P_\theta(z_T | y) + \\ 
& P_\theta(z_T| y) \nabla_\theta P_\theta(y) \log_2 P_\theta(z_T | y) 
+   P_\theta(y)\frac{\nabla_\theta P_\theta(z_T | y)}{\log 2}\Big] \\
 = & - \sum_{y \in \mathcal{O}^T} P_\theta (y) \sum_{z_T \in \{0,1\}} \Big[ \log_2 P_\theta(z_T | y) \nabla_\theta P_\theta(z_T| y) \\
& + P_\theta(z_T| y) \log_2 P_\theta(z_T | y) \nabla_\theta \log P_\theta(y) + \frac{\nabla_\theta P_\theta(z_T | y)}{\log 2} \Big].
\end{aligned}
\end{equation}

We propose a novel approach  to compute the gradients $\nabla_\theta P_\theta(z_T| y)$ and $\nabla_\theta \log P_\theta(y)$ for $z_t\in \{0,1\}$. First, we introduce the concept of forward $\alpha$ messages from \ac{hmm} \cite{baum1970maximization}. 
Given a fixed observation sequence $y = o_{[0:T]}$, for each $0\le t \le T$, the forward $\alpha$ message at the time step $t$ for a given state $j$ is,

\begin{equation}
\label{eq:forward_path_probability}
\alpha_t(j, \theta) \coloneqq P_\theta(o_0, o_1, \dots, o_t, S_t = j),
\end{equation}
which is the joint probability of receiving observation $o_{[0:t]}$ and arriving at state $j$ at the $t$-th time step.
By definition of $\alpha$ messages and the law of total probability, the probability of receiving observation $y$ is $P_\theta(y) = \sum_{i \in\mathcal{S}} \alpha_T(i, \theta)$.

The forward $\alpha$ messages can be calculated recursively by the following equations,
\begin{equation}
\label{eq:update_forward_path_probability}
\alpha_t(j, \theta) = \sum_{i \in \mathcal{S}} \alpha_{t-1}(i, \theta) P_\theta(i,j) b_j(o_t), 
\end{equation}
for $t=0,\ldots T$. The initial forward $\alpha$ message is defined jointly by the initial observation and the initial state distribution,
$\alpha_0(j,\theta) = \mu_0(j) b_j(o_0), 0 \leq j \leq N$. Since $Z_T$ is the random variable to represent if the last state $S_T $ is in a set $  E$ of secret states, then 
\begin{equation}
    P_\theta (Z_T =1, y) =  \sum_{k\in E} \alpha_T(k,\theta),
\end{equation}
and the conditional probability $P_\theta(Z_T=1| y)$ is given by
\begin{equation}
\label{eq:HMM_P_zT_y}
P_\theta(Z_T=1|y) = \sum_{k\in E} \frac{\alpha_T(k, \theta) }{P_\theta(y)}. 
\end{equation}

The derivative of $P_\theta(z_T|y)$  w.r.t. policy parameter $\theta$ can be computed as
\begin{align}
\label{eq:HMM_gradient_P_zT_y}
&\nabla_\theta P_\theta(Z_T=1 |y) = \sum_{k \in E} \nabla_\theta \frac{\alpha_T(k, \theta) }{P_\theta(y)} \nonumber\\
&= \sum_{k \in E} \frac{\nabla_\theta \alpha_T(k,\theta) P_\theta(y) - \alpha_T(k,\theta) \nabla_\theta P_\theta(y)}{ P_\theta^2(y)}\\
&= \sum_{k \in E} \Big[ \frac{\nabla_\theta \alpha_T(k,\theta)}{P_\theta(y)} - \frac{\alpha_T(k,\theta)}{ P_\theta^2(y)} \nabla_\theta P_\theta(y) \Big],  \nonumber
\end{align}
where $\nabla_\theta P_\theta(y) = \sum_{k \in \mathcal{S}} \nabla_\theta \alpha_T(k, \theta)$. 
To obtain $\nabla_\theta P_\theta(Z_T = 0|y)$, we have $P_\theta(Z_T = 0|y) = 1 - P_\theta(Z_T = 1|y)$. Thus, $\nabla_\theta P_\theta(Z_T = 0|y) = - \nabla_\theta P_\theta(Z_T = 1|y)$.

Thus, computing the gradient of $P_\theta(Z_T=1|y)$ and $P_\theta(Z_T=0|y)$ requires the gradients of all forward $\alpha$ messages w.r.t. the policy parameter $\theta$. We can compute these gradients using the following recursive computation based on  \eqref{eq:update_forward_path_probability}: For $1 \le t \le T$,
\begin{equation}
\begin{aligned}
\label{eq:update_gradient_forward_path_probability}
&\nabla_\theta \alpha_t(j, \theta) = \sum_{i=1}^N P_\theta(i,j) b_j(o_t)\nabla_\theta \alpha_{t-1}(i, \theta) \\
&+ \sum_{i=1}^N \alpha_{t-1}(i, \theta) b_j(o_t) \nabla_\theta P_\theta(i,j),
\end{aligned}
\end{equation}
and $\nabla_\theta \alpha_0(j, \theta) = 0$. It is observed that the set of forward $\alpha$ messages given the current policy $\pi_\theta$ and a fixed observation $y$ can be computed using \eqref{eq:update_forward_path_probability}. In addition, the gradient
\begin{equation}
\begin{aligned}
\label{eq:P_i_j_calculation}
\nabla_\theta P_\theta(i,j) &= \sum_{a \in \mathcal{A}} P(j|i,a) \nabla_\theta \pi_\theta(a|i)\\
&= \sum_{a \in \mathcal{A}} P(j|i,a) \pi_\theta(a|i) \nabla_\theta \log \pi_\theta(a|i)
\end{aligned}
\end{equation}
can be computed using the current policy $\pi_\theta$ and the gradient of the policy w.r.t. $\theta$. Lastly, the transition kernel $P_\theta(i,j)$ and the emission probabilities are known or can be computed. 


After computing $\nabla_\theta P_\theta(Z_T = 1 |y), \nabla_\theta P_\theta(Z_T = 0 |y)$, and $\nabla_\theta P(y)$, we can obtain the value of $\nabla_\theta H(Z_T|Y;\theta)$ by equation \eqref{eq:HMM_gradient_entropy}. It is noted that, though $\mathcal{O}^T$ is a finite set of observations, it is combinatorial and may be too large to enumerate. To mitigate this issue, we can employ sample approximations to estimate   $\nabla_\theta H_\theta(Z_T|Y)$: 
Given $M$ sequences of observations $\{y_1, \dots, y_M\}$, we can approximate $ H(Z_T|Y;\theta)$ by
\begin{equation}
\label{eq:eq:HMM_approx_entropy}
H (Z_T|Y;\theta) \approx - \frac{1}{M} \sum_{k=1}^M \sum_{z_T \in \{0,1\}} P_\theta(z_T| y_k) \log_2 P_\theta(z_T | y_k),
\end{equation}
and approximate $\nabla H(Z_T|Y;\theta)$ by
\begin{equation}
\begin{aligned}
\label{eq:HMM_approx_gradient_entropy}
&\nabla_\theta H (Z_T|Y;\theta) \\
&\approx - \frac{1}{M} \sum_{k=1}^M \sum_{z_T \in \{0,1\}} \big[ \log_2 P_\theta(z_T | y_k) \nabla_\theta P_\theta(z_T| y_k) \\
& + P_\theta(z_T| y_k) \log_2 P_\theta(z_T | y_k) \nabla_\theta \log P_\theta(y_k) + \frac{\nabla_\theta P_\theta(z_T | y_k)}{\log 2} \big].
\end{aligned}
\end{equation}

\section{Synthesizing Maximally Opacity-Enforcement Controllers For Initial-state Opacity}

In this subsection, we look into the initial-state opacity, measured by the conditional entropy $H(S_0|Y;\theta)$ of the initial state $S_0$ given by the observation sequence $Y$, which is defined by:
\begin{equation}
\label{eq:HMM_entropy_2}
H(S_0|Y;\theta) = - \sum_{y \in \mathcal{O}^T} \sum_{s_0 \in \mathcal{S}} P_\theta(s_0, y) \log_2 P_\theta(s_0 | y). 
\end{equation}

The maximal opacity-enforcement planning for initial-state opacity can be formulated similarly as a constrained optimization problem:
\begin{equation}
\begin{aligned}
\label{eq:back_opt_problem_1}
& \max_\pi && H (S_0|Y;\theta) \\
&\optst: && V(s_0,\theta) \geq \delta,
\end{aligned} 
\end{equation}
where $s_0\in \mathcal{S}$ is the initial state sampled from $\mu_0$. In this optimization problem, we assume that $s_0$ is known to the planning agent but not the observer.  

For this constrained optimization problem with an inequality constraint, we formulate the problem \eqref{eq:back_opt_problem_1} into the following max-min problem for the associated Lagrangian $L (\theta, \lambda)$:
\begin{equation}
\label{eq:Lagrangian_2}
\max_\theta \min_{\lambda \geq 0} L(\theta, \lambda) = H(S_0|Y;\theta) + \lambda (V(s_0, \theta) - \delta),
\end{equation}
where $\lambda$ is the multiplier. 

Following the similar primal-dual optimization approach (see \eqref{eq:primal_dual_algorithm}),  we need to calculate the gradient of $H(S_0|Y;\theta)$ w.r.t. the policy parameter. Also, using a trick that $\nabla_\theta P(y) = P_\theta(y) \nabla_\theta \log P(y)$, we have
\begin{equation}
\begin{aligned}
\label{eq:HMM_gradient_posterior_entropy}
&\nabla_\theta H(S_0|Y;\theta) \\ 
&= - \nabla_\theta \sum_{y \in \mathcal{O}^T} P_\theta(y) \sum_{s_0 \in \mathcal{S}} P_\theta(s_0| y) \log_2 P_\theta(s_0 | y)\\
&= - \sum_{y \in \mathcal{O}^T} P_\theta (y) \sum_{s_0 \in \mathcal{S}} \Big[ \log_2 P_\theta(s_0 | y) \nabla_\theta P_\theta(s_0| y) \\
&+ P_\theta(s_0| y) \log_2 P_\theta(s_0 | y) \nabla_\theta \log P_\theta(y) + \frac{\nabla_\theta P_\theta(s_0 | y)}{\log 2}  \Big]. 
\end{aligned}
\end{equation}
The computations of  $P_\theta(\theta)$ and $\nabla_\theta \log P_\theta(y)$ are the same as those for the last-state opacity (see Section~\ref{subsec:forward_gradient}). In addition, for the initial-state opacity, we need to obtain the $P_\theta(s_0|y)$ and its derivative w.r.t. $\theta$.


From Bayes' theorem, 
\begin{equation}
\label{eq:bayes_rule}
P_\theta(s_0|y) = \frac{P_\theta(y|s_0) \mu_0(s_0)}{P_\theta(y)}.
\end{equation}
Note that $\mu_0$ is the prior distribution of the initial state, which is known and does not depend on $\theta$. 
Thus, the gradient of $P_\theta(s_0|y)$ w.r.t. $\theta$ is given by
\begin{equation}
\begin{aligned}
\label{eq:gradient_bayes_rule}
\nabla_\theta P_\theta(s_0|y) =  \frac{\mu_0(s_0)}{P_\theta(y)} \nabla_\theta P_\theta(y|s_0) \\
- \frac{\mu_0(s_0) P_\theta(y|s_0)}{P_\theta^2(y)} \nabla_\theta P_\theta(y).
\end{aligned}
\end{equation}

To compute the $P_\theta(y| s_0)$ and $\nabla_\theta P_\theta(y| s_0)$ for $s_0 \in \mathcal{S}$, we employ the backward $\beta$ messages from HMM \cite{baum1970maximization}, defined as follows: 
Given a observation sequence $o_{[t:T]}$, for each $0\le t \le T$, the backward $\beta$ message at the time step $t$ for a given state $i$ is,
\begin{equation}
\label{eq:backward_definition}
\beta_t(i,\theta) \coloneqq P(o_t,o_{t+1},\dots,o_T|S_t = i),
\end{equation}
which represents the probability of having the observation sequence $o_{[t:T]}$ given the state is $i$ at time step $t$. It can be computed by the backward algorithm:
\begin{equation}
\label{eq:backward_algorithm}
\beta_t(i, \theta) = \sum_{j \in \mathcal{S}} \beta_{t+1}(j, \theta) P_\theta(i,j) b_j(o_{t+1}),
\end{equation}
for $ t=0,\ldots T-1$, and $\beta_T(i,\theta) = 1$ for $1 \leq i \leq N$. Also
note that $P_\theta(y|i) = \beta_0(i,\theta)$ for each state in the support of the initial state distribution.  
By recursion, we can obtain $P_\theta(y|i)$ using the backward $\beta$ messages. For each $i\in \supp(\mu_0)$, we can also obtain the gradient of $P_\theta(y|i)$   by
\begin{multline}
\label{eq:gradient_backward_algorithm}
 \nabla_\theta P_\theta(y|i) = \nabla_\theta \beta_0(i, \theta) =\\
\sum_{j \in \mathcal{S}} [ P_\theta(i,j) b_j(o_{t+1}) \nabla_\theta \beta_{t+1}(j, \theta) \\
 +  b_j(o_{t+1}) \beta_{t+1}(j, \theta) \nabla_\theta  P_\theta(i,j)]. 
\end{multline}
The computation of $\nabla_\theta P_\theta(i,j)$ is in equation~\eqref{eq:P_i_j_calculation}. In this way, $H(S_0|Y;\theta)$ and $\nabla_\theta H(S_0|Y;\theta)$ can be computed exactly. Again, since $y \in \mathcal{O}^T$ is combinatorially many, we employ sample approximation to estimate the value. Given $M$ trajectories $\{y_1, \dots, y_M\}$, the entropy term $H(S_0|Y;\theta)$ and the gradient term $\nabla_\theta H(S_0|Y;\theta)$ can be approximated by
\begin{equation*}
\label{eq:eq:backward_approx_entropy}
H (S_0|Y;\theta) \approx - \frac{1}{M} \sum_{k=1}^M \sum_{s_0 \in \mathcal{S}} P_\theta(s_0| y_k) \log_2 P_\theta(s_0| y_k),
\end{equation*}
and
\begin{align*}
\label{eq:backward_approx_gradient_entropy}
&\nabla_\theta H (S_0|Y;\theta) \approx - \frac{1}{M} \sum_{k=1}^M \sum_{s_0 \in \mathcal{S}} [ \log_2 P_\theta(s_0 | y_k) \nabla_\theta P_\theta(s_0| y_k) \\
& + P_\theta(s_0| y_k) \log_2 P_\theta(s_0 | y_k) \nabla_\theta \log P_\theta(y_k) + \frac{\nabla_\theta P_\theta(s_0 | y_k)}{\log 2} ]. 
\end{align*}

With the above step of calculating the gradient of conditional entropy $H (S_0|Y;\theta)$ w.r.t. $\theta$, we can then employ the primal-dual approach to solve a (local) optimal solution to Problem~\ref{problem:initial-state}.

\section{Experiment Evaluation}

\begin{example}[Grid World Example]
The effectiveness of the proposed optimal opacity-enforcement planning algorithms \footnote{The code of the experiment is available on \url{https://github.com/AronYoung414/leakage_minial_design_MDP}} is illustrated through a stochastic grid world example shown in Figure~\ref{fig:grid_world_env}. In this example, we focus on optimizing the last-state opacity and initial-state opacity. The details of the environment setting are outlined in Figure~\ref{fig:grid_world_env}. For perception, four sensors are placed on the grid with distinct ranges indicated by the blue, red, yellow, and green areas in the picture. As P1 enters these sensor ranges, the observer receives corresponding observations (``b", ``r", ``y", ``g", respectively) with probability $p=0.9$ and a null observation (``0") with probability $1-p=0.1$, 
attributed to the false negative rate of the sensors.
\begin{figure}[tp!]
\includegraphics[width=0.6\linewidth]{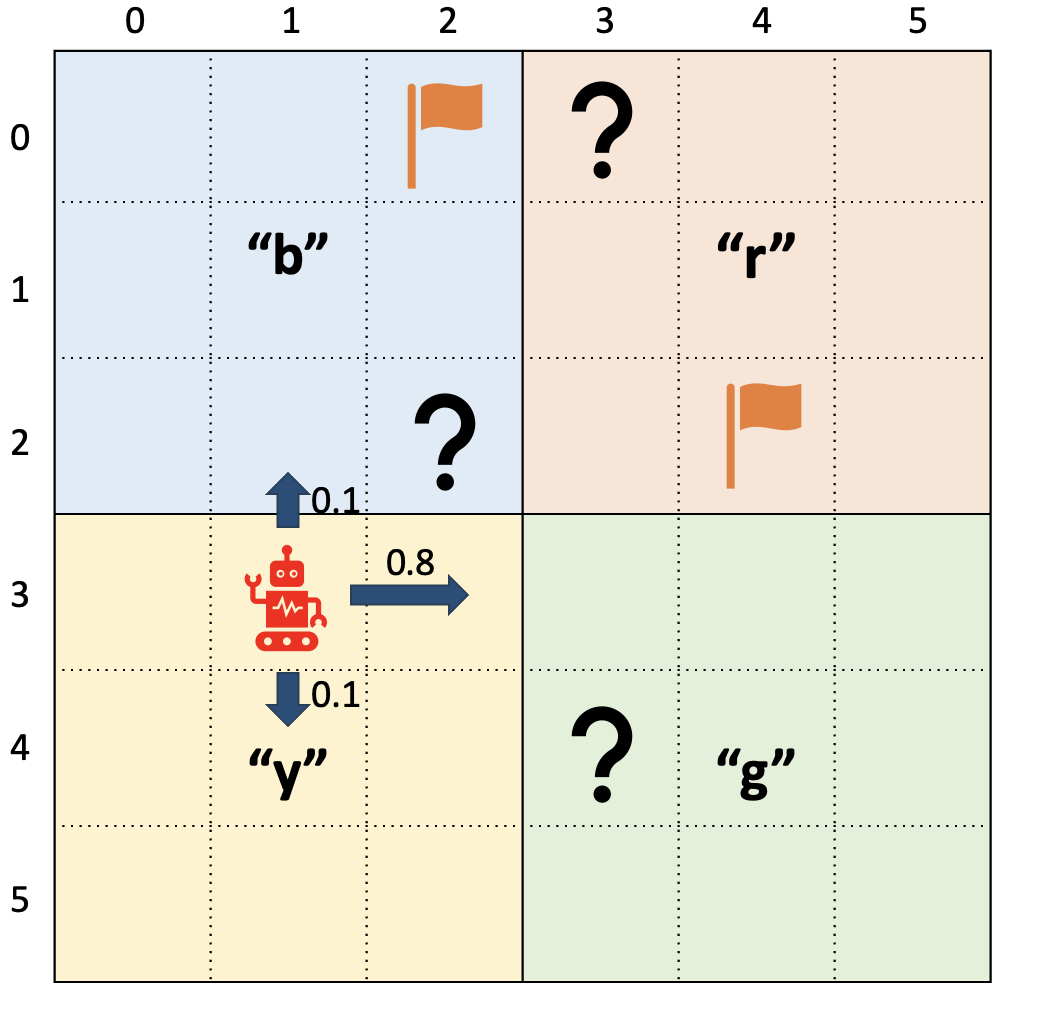}
\centering
\caption{The red robot is P1 (the agent). P1 can move in four compass directions (north, south, east, west) or remain stationary. However, the dynamics of movement are stochastic. When the robot moves in a specific direction, there is a 0.1 probability that it will also move in the nearest two directions. For instance, if the robot moves east, there is a 0.1 probability of it moving north and a 0.1 probability of it moving south, as illustrated in the image. If the robot hits the boundary, it stays put. }
\label{fig:grid_world_env}
\end{figure}
\end{example}

The question marks on the grid represent the secret states for P1, while the flags denote the goal states for P1.  We set the reward of reaching a goal to be 0.1 and the constraint that the total return is greater than or equals $\delta = 0.3$, and the horizon $T = 10$. Note that, in this finite horizon, the robot can repeatedly visit the goal state.

We will employ the soft-max policy parameterization, i.e.,
\begin{equation}
    \pi_{\theta}(a|s) = \frac{\exp(\theta_{s,a})}{\sum_{a' \in \mathcal{A}}\exp(\theta_{s, a'})},
\end{equation}
where $\theta\in \reals^{
|S\times A|}$ is the policy parameter vector.
The softmax policy has good analytical properties including completeness and differentiability.

\begin{figure*}[htp!]
\centering
\begin{subfigure}{.5\textwidth}
  \centering
  \includegraphics[width=0.7\linewidth]{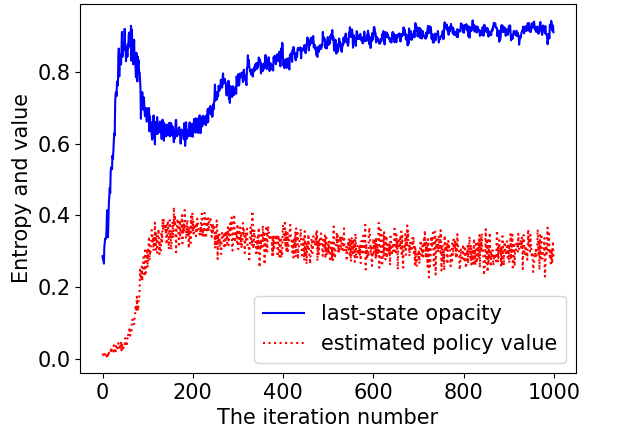}
  \caption{Last-state opacity}
  \label{fig:last_opacity_results}
\end{subfigure}\hfill
\begin{subfigure}{.5\textwidth}
  \centering
  \includegraphics[width=0.7\linewidth]{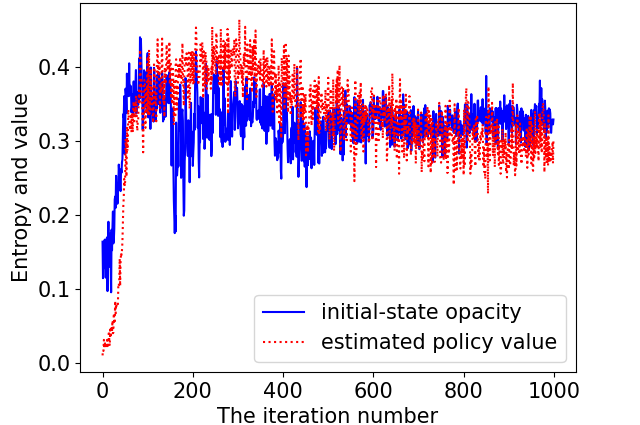}
  \caption{Initial-state opacity}
  \label{fig:initial_opacity_results}
\end{subfigure}
\caption{The result of the primal-dual policy gradient algorithm. The blue line represents the opacity and the red line represents the estimated total return.}
\label{fig:primal_dual_results}
\end{figure*}

During the optimization of last-state opacity, we maintain a fixed initial state for P1, given by the position of the robot illustrated in Figure~\ref{fig:grid_world_env}. Figure~\ref{fig:last_opacity_results} illustrates the estimated value of the last-state opacity and the total return within the primal-dual policy gradient method. We use the estimated total return instead of the analytical total return from value iterations because we employ the gradient estimate of the total return in the proposed opacity-enforcement policy gradient method.

\begin{figure*}[ht!]
\centering
\begin{subfigure}{.33\textwidth}
  \centering
  \includegraphics[width=\linewidth]{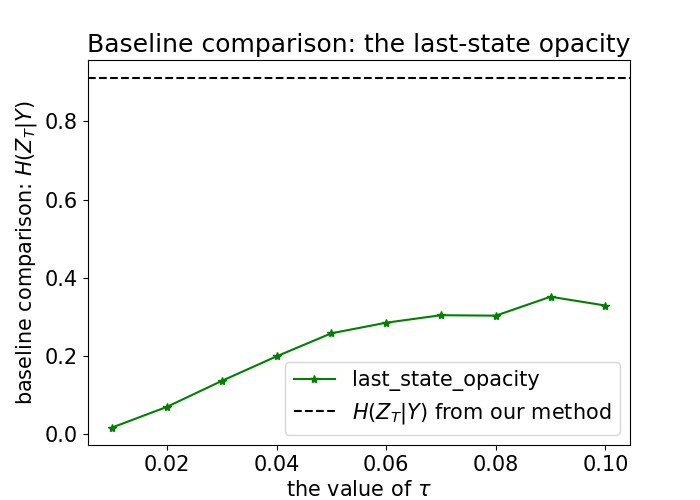}
  \caption{Last-state opacity}
  \label{fig:baseline_last_entropy}
\end{subfigure}%
\begin{subfigure}{.33\textwidth}
  \centering
  \includegraphics[width=\linewidth]{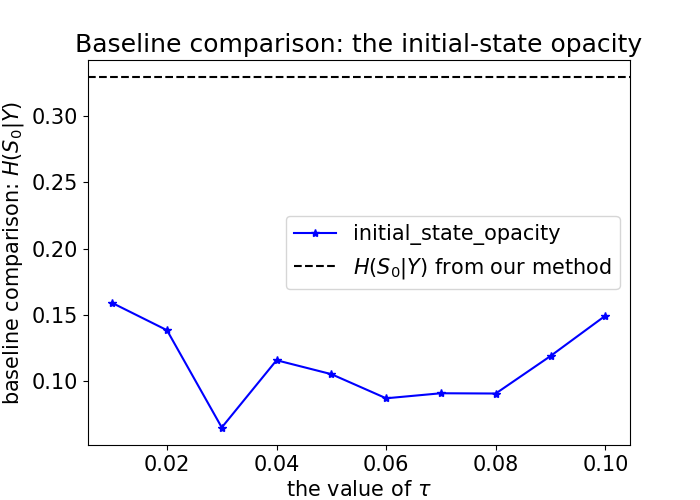}
  \caption{Initial-state opacity}
  \label{fig:baseline_initial_entropy}
\end{subfigure}
\begin{subfigure}{.33\textwidth}
  \centering
  \includegraphics[width=\linewidth]{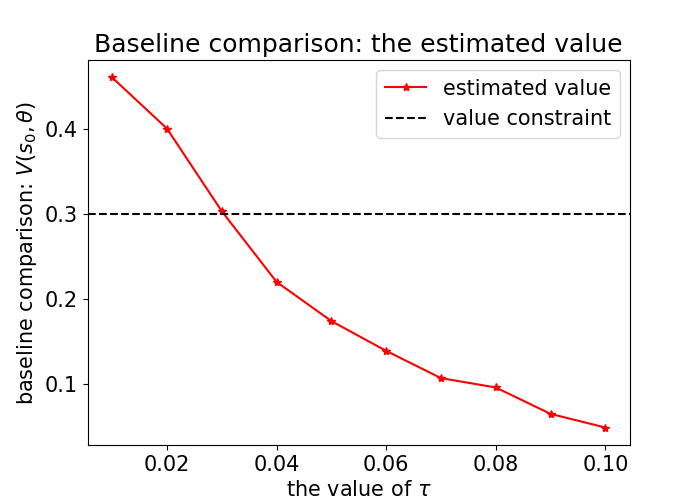}
  \caption{Value $V(S_0,\theta)$ of the policy}
  \label{fig:baseline_value}
\end{subfigure}
\caption{Comparison with baseline. The dashed lines are the entropies and values from our method. The colorful lines are entropies and values from the baseline method.}
\label{fig:baseline}
\end{figure*}

Figure~\ref{fig:last_opacity_results}   illustrates when the algorithm converges, the conditional entropy $H(Z_T|Y)$ eventually approaches $0.930$. The value of policy reaches $0.328$, satisfying the predefined threshold of $\delta = 0.3$. 
Given the conditional entropy is close to 1, which is the maximal value of the entropy, the observation reveals little information regarding whether a secret location is visited, even when the observer knows the exact policy used by the robot.
The conditional entropy is inversely proportional to the value/total return. This behavior is a consequence of the environmental configuration. From observing the sampled trajectories, we noticed that the agent navigates among various sensor ranges to confuse the observer, thereby diminishing the total return that P1 can achieve, as P1 is constrained from remaining stationary at the goal.

During the optimization of initial-state opacity, the set of possible initial states is four corners (cells $(0,0)$, $(0,5)$, $(5,0)$, $(5,5)$). The initial-state distribution follows a discrete uniform distribution across this set of initial states.  Figure~\ref{fig:initial_opacity_results} illustrates the initial-state opacity, measured by the conditional entropy $H(S_0|Y)$, and the value of the policy within the primal-dual policy gradient method. The same threshold $\delta = 0.3$ is employed here.
When the algorithm converges, the initial conditional entropy eventually reaches $0.329$. The policy value reaches $0.301$, satisfying the predefined threshold of $\delta = 0.3$. In this environment, elevated initial-state opacity indicates the observer is less uncertain about P1's specific initial state. This result is understandable because the four sensors are highly sensitive to detect the presence of P1 in its range and different initial states give different sensor readings to the observer with high probabilities.

Since there are no other algorithms for solving the proposed opacity-enforcement planning problems, we consider a comparison with a baseline algorithm for entropy-regularized \ac{mdp}s, where the objective value is a weighted sum of the total return and the discounted entropy of the policy \cite{Nachum2017bridge}. As the weight on the entropy terms increases, the policy is more stochastic and noisy.

The objective function for entropy-regularized \ac{mdp}s is given by  \cite{Nachum2017bridge,Cen2022fastentropy}:

\begin{equation}
V_\tau(s;\theta) = V(s;\theta) + \tau H(s;\theta).
\end{equation}
The discounted entropy term, $H(s;\theta)$, is defined as:
\begin{equation*}
\begin{aligned}
& H(s;\theta) = -\frac{1}{1 - \gamma} E_{s \sim d_{\pi_\theta}} \\
&\Big[ \sum_{a \in \mathcal{A}} \pi_\theta(a|s) \log \pi_\theta(a|s) \Big],
\end{aligned}
\end{equation*}
where $d_{\pi_\theta}$ is the occupancy measure induced by policy $\pi_\theta$. The objective is to maximize the regularized total return $V_\tau(s;\theta)$.


We compare our method with entropy-regularized MDPs solved with different $\tau$ values, selecting $10$ values from $\tau = 0.01$ to $\tau = 0.1$. The following graphs (Figure~\ref {fig:baseline}) illustrate the results. Note that, when $\tau = 0.1$, the policy approaches a random policy, making it impractical to further increase opacity by raising $\tau$.


The results in Figure~\ref {fig:baseline} indicate that maximizing last-state opacity and initial-state opacity cannot be achieved using standard entropy-regularized MDP. 
In Figure~\ref{fig:baseline_last_entropy}, we observed that, despite the increasing value for $\tau$, the conditional entropy $H(Z_T|Y)$ under the baseline algorithm remains below 0.4 and below the value obtained with the proposed method, indicating that the observer has more certainty in predicting if a secret state is visited. The value of the policy decreases as $\tau$ increases, and when $\tau \ge 0.03$, the entropy-regulated policy fails to satisfy the constraint on the total return. Our method outperforms this baseline by attaining the value of $0.3$ with a last-state opacity of $0.930$.  
Figure~\ref{fig:baseline_initial_entropy}
shows that the initial-state opacity fluctuates with increasing $\tau$ but does not reach the optimal initial-state opacity  (0.329) achieved by our method.

These comparison experiments indicate that our proposed methods could exploit the noise in the observation to optimize opacity. This feature is not achievable using a policy computed from entropy-regularized \ac{mdp}s, which do not employ the observation function.  
\section{Conclusion and Future Work}
 
In this paper, we introduce information-theoretic definitions of opacity in dynamical systems modeled as Markov decision processes, against observers with imperfect information. Given the constraints on task performance, we develop opacity-enforcement policy gradient methods based on message passing in Hidden Markov Models (HMMs) for two basic types of opacity: last-state opacity and initial-state opacity.
We expect the fundamental results on last-state opacity and initial-state opacity would be extended to other types of opacity, including language-based and model-based opacity and other obfuscation methods.


While the algorithm exhibits good performance in experiments, its drawback lies in high computational complexity due to the utilization of the forward-backward algorithm for gradient calculation. Finding alternative approaches to mitigate computation complexity will be among our future steps. 
Another direction to explore would be to investigate opacity-enforcement under different assumptions for the observer, for example, observers with imprecise knowledge about the agent's policy. Lastly, though the objective is to maximize opacity, the same approaches can be used for minimizing opacity, and thus maximizing transparency, which is a desired property for AI and robotic systems interacting with humans. Future work would also look into potential applications and evaluate the effects of transparency in multi-agent teaming.




\section*{Acknowledgements}

This work was sponsored in part by Army Research Laboratory under Cooperative Agreement Number W911NF-22-2-0233,  the Army Research Office and was accomplished under Grant
Number W911NF-22-1-0166
 and   in part by NSF under grant No. 2144113. The views and conclusions contained in this document are those of the authors and
should not be interpreted as representing the official policies, either expressed or implied, of the Army Research
Office, the Army Research Lab, or the U.S. Government. The U.S. Government is authorized to reproduce and distribute reprints for
Government purposes notwithstanding any copyright notation herein.
\bibliographystyle{named}
\bibliography{ijcai24}

\end{document}